\documentstyle[epsfig]{aipproc}
\begin{document}
\title{Where is the COBE maps' non-Gaussianity?}
\author{Jo\~{a}o Magueijo$^1$, 
Pedro G. Ferreira$^{2,3}1$, and  Krzysztof M. 
G\'orski$^{4,5}$}

\address{
$^1$Theoretical Physics, 
Imperial College, Prince Consort Road, London SW7 2BZ, UK\\
$^2$ Theory Group, CERN, CH-1211, Geneve 23 ,Switzerland\\
$^3$ CENTRA, Instituto Superior Tecnico, Lisbon, Portugal\\
$^4$Theoretical 
Astrophysics Center, Juliane Maries Vej 30,
DK-2100, Copenhagen \O, Denmark\\
$^5$ Warsaw University Observatory, Warsaw, Poland}
\maketitle
\begin{abstract}
We review our recent claim that there is evidence of non-Gaussianity
in the 4 Year COBE DMR data. We present some new results concerning the effect
of the galactic cut upon the non-Gaussian signal. These findings
imply a localization of the non-Gaussian signal on the Northern
galactic hemisphere.
\end{abstract}

\section{Evidence for CMB non-Gaussianity}
In a recent paper \cite{apjl1} we showed that the 4 Year COBE DMR data
exhibits evidence of non Gaussianity at a high confidence level. 
We made use of statistical tools described in more detail in 
\cite{santa,bigpaper}. Since then our result 
has been corroborated by two other groups
\cite{nov,pando}. 
In this review we revisit our analysis, and the tests to which we have
subjected it, including some new results.

In our analysis we propose, and work with, an estimator for
the {\it normalized bispectrum}, denoted by $I^3_\ell$.
We refer the reader to \cite{apjl1} for its definition. We then
applied this estimator to 
the  inverse noise variance weighted, average maps of 
the 53A, 53B, 90A and 90B {\it COBE}-DMR channels, with monopole
and dipole removed, at resolution 6, in ecliptic 
pixelization. We use the  
extended galactic cut of \cite{banday97}, and 
\cite{benn96} to remove most of the emission from the plane of the Galaxy.
We apply our statistics to the DMR maps before and after correction
for the plausible diffuse foreground emission outside the galactic plane
as described in
\cite{kog96b}, and \cite{COBE}.

\begin{figure}
\centering
\leavevmode\epsfysize=7cm \epsfbox{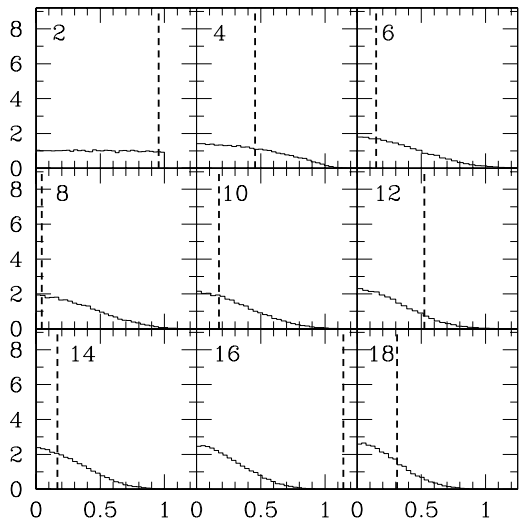}\\ 
\caption[flatness]{\label{fig1}The vertical thick dashed line represents the value 
of the observed
$I^3_\ell$.  The solid line is the probability distribution function
of $I^3_\ell$ for a Gaussian sky with extended galactic cut and
DMR noise.}
\end{figure}

By means of Monte Carlo simulations we also 
found the distributions $P_\ell(I^3_\ell)$ for what
we should have seen assuming a Gaussian signal, which is then
processed by the experimental set up associated with DMR. 
These $P(I^3_\ell)$ were inferred 
from 25000 realizations (see Fig.~\ref{fig1}). 
The 
observed $I^3_\ell$ and the distributions $P_\ell(I^3_\ell)$
are plotted in Fig.~\ref{fig1}. One immediately notices the presence
of a significant deviant.

In order to quantify this deviant we define the goodness of fit
statistic
\begin{equation}\label{presc}
X^2={1\over N}{\sum_\ell X_\ell^2}=
{1\over N}{\sum_\ell (-2\log P_\ell(I^3_\ell) 
+ \beta_\ell),}
\end{equation}
where the constants $\beta_\ell$ are defined so that for each term
of the sum $\langle X_\ell^2\rangle=1$. The definition reduces
to the usual chi squared for Gaussian $P_\ell$. 
We build a $X^2$ for the {\it COBE}-DMR data from the 
$P_\ell(I^3_\ell)$ inferred from Monte Carlo simulations, 
taking special care with the numerical
evaluation of the constants $\beta_\ell$. We call 
this function $X^2_{COBE}$.
We then find its distribution $F(X^2_{COBE})$
from 10000 random realizations.  This is very well approximated by 
a $\chi^2$ distribution with 12 degrees of freedom.
We then compute $X^2_{COBE}$ with the actual observations and find
$X^2_{COBE}=1.81$. One can compute $P(X^2_{COBE}<1.81)= 0.98$.
Hence, it would appear that we can
reject Gaussianity at the $98\%$ confidence level.

\section{Is it a systematic effect?}
We checked that this result could not be due to the following
systematics:
\begin{quote}
\begin{enumerate}
\item Foregrounds contamination:
\begin{itemize}
\item Dust (using the DIRBE sky maps and also the Schlegel {\it et al}
dust model)
\item Synchrotron (with the Haslam template)
\item Foreground corrected maps (effect persists on corrected
maps)

\end{itemize} 
\item Noise model:
\begin{itemize}
\item Anisotropic sky coverage
\item Noise correlations between different pixels 
\item Analysis of noise templates
\end{itemize}
\item Galactic cut:
\begin{itemize}
\item Dependence on shape (``custom'' versus constant elevation)
\item Dependence on elevation
\item Dependence on monopole and dipole subtraction, before or
after the cut, with or with out galaxy.
\end{itemize}
\item  
Possible small residual errors in corrections for  
\begin{itemize}
\item Spurious offsets induced by the cut.
\item Instrument susceptibility to the Earth 
magnetic field.
\item Callibration errors .
\item Errors due to incorrect removal of the COBE Doppler and  
Earth Doppler signals.
\item Errors in correcting for  emissions from the Earth, and
eclipse effects.
\item Artifacts due to uncertainty in the correction for
the correlation created by the low-pass filter on the
lock-in amplifiers (LIA) on each radiometer
\item Errors due to emissions from the moon, and the planets. 
\end{itemize}
\item Assumptions in Monte Carlos:
\begin{itemize}
\item Dependence on power spectrum tilt
\item Dependence on smooth versus discontinuous power spectrum
\item Dependence on beam shape
\item Dependence on pixelization.
\end{itemize}
\end{enumerate}
\end{quote}

In fact, the confidence level quoted above reflects the worse line up of
systematics. If we try to correct for systematics, in general the confidence
level for rejecting Gaussianity is enlarged to beyond 99\%, as we
describe in more detail in \cite{bigpaper}

\section{Where is the non-Gaussianity?}
\label{cuts}
We now concentrate on a subset of tests involving the galactic cut
which we applied to our result. 
Changing the galactic cut affects sample variance besides 
eliminating possible contaminations from the map. We considered
extensive variations of the cut, including additions of  polar cap cuts 
to the extended cut. 
\begin{figure}
\centerline{\psfig{file=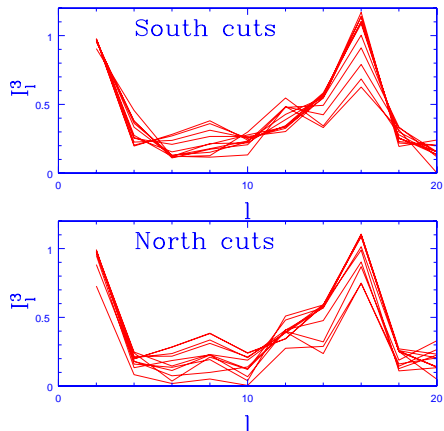,width=6cm}}
\caption{The effect of combining the extended cut with polar cap cuts down
to $60^\circ$ in the North or South hemispheres. }
\label{cutns}
\end{figure}

We found that cuts from the pole affect the result more than cuts from
the equator. This suggests that the effect may be localized near the Poles.
We therefore decided to compare the effect of applying cuts only in the North or
South galactic poles. We considered cuts down to $60^\circ$
(2668 pixels excluded). We find the curious result that cutting Northern
caps is more damaging for the nonGaussian spike than cutting
Southern caps (fig.~\ref{cutns}). 
Indeed the first few Southern cap cuts appear to increase
the spike.

Non-Gaussianity could in principle be localized in Fourier
space without being localized 
in real space\footnote{See New Scientist, Ed comment
to letter, 12/12/98, for a layman's opinion.}. Examples of such behaviour 
are given 
in \cite{fermag}. We believe that the signal we have found is 
essentially localized in Fourier space.
However the results we have just presented suggest that 
our signal may indeed be also localised in real space, 
around the Northern galactic
cap.
\\

{\bf Acknowledgements}
JM would like to thank the organizers for an excellent meeting. We thank
JNICT, NASA-ADP, NASA-COMBAT, NSF,  RS, Starlink, TAC for support.

\end{document}